\documentclass[aps,prmaterials,reprint]{revtex4-2}

\pdfoutput=1
\usepackage{graphics,epsfig}
\usepackage{amsmath,amssymb}
\usepackage{bm,color}
\usepackage{subfigure}

\begin{document}

\title{Atomic ordering and phase separation in lateral heterostructures and multijunctions of
  ternary two-dimensional hexagonal materials}
\author{Zhi-Feng Huang}
\affiliation{Department of Physics and Astronomy, Wayne State University,
Detroit, Michigan 48201, USA}

\date{\today}

\begin{abstract}
The growth and microstructural properties of ternary monolayers of two-dimensional hexagonal
materials are examined, including both individual two-dimensional crystalline grains and
in-plane heterostructures, multijunctions, or superlattices. The study is conducted through
the development of a ternary phase field crystal model incorporating sublattice ordering and
the coupling among the three atomic components. The results demonstrate that a transition of
compositional pattern or modulation in this type of two-dimensional ternary crystals, from
phase separation to geometrically frustrated lattice atomic ordering, can be controlled
via the varying degree of energetic preference of heteroelemental neighboring over the
homoelemental ones. Effects of growth and system conditions are quantitatively identified
through numerical calculations and analyses of interspecies spatial correlations and the
degree of alloy intermixing or disordering. These findings are applied to simulating the
growth of monolayer lateral heterostructures with atomically sharp heterointerface, and via
the sequential process of edge-epitaxy, the formation of the corresponding superlattices or
structures with multiple heterojunctions, with outcomes consistent with recent experiments
of in-plane multi-heterostructures of transition metal dichalcogenides. Also explored is
a distinct type of alloy-based lateral heterostructures and multijunctions which integrate
ternary ordered alloy domains with the adjoining blocks of binary compounds, providing a
more extensive variety of two-dimensional heterostructural materials.

\end{abstract}

\maketitle

\section{Introduction}

While the two-dimensional (2D) edge-epitaxial growth of in-plane lateral heterostructures
has been achieved recently, including the fabrication of heterojunctions connecting different
types of monolayer blocks of transition metal dichalcogenides (TMDs) \cite{HuangNatMater14,
GongNatMater14,DuanNatNanotech14,LiScience15,ChenAdvMater15,BogaertNanoLett16,ZhangAdvMater18,
ZhangAdvMater19} or between graphene and hexagonal boron nitride (\textit{h}-BN) \cite{LiSmall16},
and the \textit{in situ} synthesis of various TMD lateral multijunctions or superlattices
\cite{ZhangScience17,SahooNature18,XieScience18,ChiuAdvMater18,KobayashiACSNano19},
their further development is hindered by the difficulties
related to the controllability of the growth process and heterointerfacial properties.
The major challenges in controlling the growth and property of these heterostructures involve
various coupled factors, such as the structural control at the atomic level, concerning the
sharpness of heterointerfaces and defect formation, and the compositional control with effects
of intermixing or alloying of different atomic components either at interface or in bulk.
They are key to determining the functionality of heterostructures and their device applications,
and to enabling the engineering of electronic band structures, thermal transport, or magnetic
property, with examples including the fabrication of TMD-based p-n junctions
\cite{HuangNatMater14,GongNatMater14,DuanNatNanotech14,LiScience15} or metal-semiconductor
heterojunctions \cite{ZhangAdvMater18}, metal-insulator (e.g., graphene-\textit{h}-BN) lateral
tunneling structures \cite{LiSmall16}, and atomically thin in-plane quantum wells and quasi-1D
superlattices \cite{ZhangScience17,SahooNature18,XieScience18}.

Another key factor involved, which is intrinsic to multicomponent material systems, is
related to the behavior of mixing vs demixing between different atomic species, yielding
the phenomenon of short-range lattice atomic ordering vs phase separation. For an $AB$/$CB$
type heterostructure, such as TMD/TMD heterostructure in the form of MX$_2$/M$'$X$_2$
or MX$_2$/MX$'_2$ with M, M$'$ a transition metal element (e.g., Mo, W, Nb, Re) and
X, X$'$ a chalcogen element (e.g., S, Se, Te), phase separation between domains
of $AB$ and $CB$ compounds seemingly occurs. However, the 2D bulk state of $A_xC_{1-x}B_2$
ternary TMD alloys, such as Mo$_{1-x}$W$_x$X$_2$ \cite{GanSciRep14,YangChemMater18} and
Re$_{0.5}$Nb$_{0.5}$S$_2$ \cite{AziziPRL20}, does not reveal any phase-separated behavior;
instead, either random alloying or atomic ordering between $A$ and $C$ components (which
could form a triangular lattice with geometric frustration \cite{AziziPRL20}) has been found.
This can be attributed to the energetic competition or preference between heteroelemental
($A$-$C$) and homoelemental ($A$-$A$ and $C$-$C$) interactions, which in 2D TMD alloys
gives the energy gain for forming $A$-$C$ heteroatomic neighboring \cite{GanSciRep14} as
a result of charge transfer between different species \cite{YangChemMater18}. Thus, to
synthesize a lateral heterostructure or multijunction separating $AB$ and $CB$ phases via
e.g., chemical or physical vapor deposition, instead of a 2D bulk growth of ternary mixture,
experimentally a sequential process of edge-epitaxy through the control of deposition flux
is used, with the next block of new material (e.g., $CB$) grown epitaxially from the edge
of the as-grown domain of a different material (e.g., $AB$). Repeating this sequence leads
to the formation of in-plane multijunctions or superlattices with alternating domains or
blocks of different 2D materials \cite{ZhangScience17,SahooNature18,XieScience18}.

The above structural and compositional properties of 2D materials can be varied through
the control of sample growth conditions, with a typical one being the growth temperature.
For the bulk state of ternary TMD alloys, high enough temperature would result in the
disordering and random alloying of the monolayer material such as Mo$_{1-x}$W$_x$S$_2$
\cite{ChenACSNano13,GanSciRep14}. Variation of growth temperature also affects the interfacial
sharpness of heterostructural systems. Enhanced diffusion at higher temperature (and low
growth rates) leads to more compositionally intermixed and diffuse or roughening interfaces
with substitutional alloying, while at low temperatures (and/or fast deposition rates with
limited edge-diffusion process) narrow or atomically sharp heterointerfaces are generated,
as observed in experiments of in-plane TMD/TMD heterojunctions
\cite{HuangNatMater14,DuanNatNanotech14,BogaertNanoLett16,ChiuAdvMater18}.

Although both experimental and theoretical efforts have been devoted to examining those various
factors described above, the understanding of them is still far from complete due to the complexity
of the dynamical growth process of heterostructures and the large spatial and temporal scales
involved. Most of the related theoretical analyses and modeling are based on small-system atomistic
calculations [including first-principles calculations and molecular dynamics (MD) or Monte Carlo
simulations] with limited time and length scales, while a systematic understanding is still
lacking. What is also lacking, in both experimental and theoretical or computational works,
is how to incorporate the effect of atomic ordering of ternary 2D materials into heterostructures
or multijunctions, which is expected to induce more tunable functional property of the
heterostructural system particularly given the change of electronic property as a result of
atomic ordering in each individual domain of ternary 2D alloy \cite{AziziPRL20}. The corresponding
growth process will be explored in this work through the development of a phase field crystal
(PFC) model for describing the structure and dynamics of in-plane $AB$/$CB$ type ternary hexagonal
materials. 

The ternary density-field model introduced here is based on the PFC approach which is able to
resolve microscopic crystalline details of the material system and simulate dynamical
processes at large diffusive timescales
\cite{ElderPRB07,Huang08,GreenwoodPRL10,Mkhonta13,EmmerichAdvPhys12}. This method has been used
in a wide range of applications such as the study of binary and multicomponent alloy systems
\cite{Ofori-OpokuPRB13,AlsterPRE17,AlsterPRMater17,Salvalaglio21} and 2D hexagonal materials
\cite{HirvonenPRB16,SeymourPRB16,Taha17,Taha19,Zhou19,HirvonenPRB19,ElderPRM21,DongPCCP18,Waters22}.
The ternary PFC model developed in this work is for modeling the mixture or alloying
of $AB$ and $CB$ type 2D material compounds, and is applied to examining the properties of 2D
monolayers subjected to vertical confinement when deposited on a substrate (as in the epitaxial
growth of real materials). These include the control of phase separation vs atomic ordering
in a 2D ternary grain, the characterization of spatial correlation and composition intermixing
or disordering under different system conditions, and the formation of laterally edge-epitaxial
$AB/CB$ heterostructures and multijunctions consisting of alternating $AB$ and $CB$
blocks with heterointerfaces along the zigzag crystalline direction, which are consistent
with recent experimental findings of MoX$_2$/WX$_2$ (X = S, Se) and MS$_2$/MSe$_2$ (M = Mo, W)
2D heterostructural materials synthesized via multi-step, sequential epitaxial growth.
The extension to incorporate atomically ordered $ACB$-type ternary alloy blocks into an
interesting new kind of alloy-based in-plane multi-heterostructures is predicted, as
demonstrated in our PFC simulations generating lateral heterojunctions comprising adjacent
blocks of 2D binary compounds and ternary ordered alloys.

\section{Model}
\label{sec:model}

The PFC model for binary $AB$ sublattice ordering can be developed from classical dynamical
density functional theory (DDFT) \cite{Taha17,Taha19}, where the model system is described
by the evolution of atomic density variation fields $n_\eta$ ($\eta=A,B$) for $A$ and $B$
components, i.e.,
\begin{equation}
  \frac{\partial n_\eta}{\partial t} = m_\eta \nabla^2 \frac{\delta \mathcal{F}}{\delta n_\eta}.
  \label{eq:n_dynamics}
\end{equation}
Here $m_A=1$ after rescaling, $m_B=M_B/M_A$ with $M_A$ and $M_B$ the atomic mobilities of
$A$ and $B$ species, and the rescaled PFC free energy functional $\mathcal{F} = \mathcal{F}_A
+ \mathcal{F}_B + \mathcal{F}_{AB}$, where $\mathcal{F}_{\eta=A,B}$ is the same as that of
single-component PFC, i.e.,
\begin{eqnarray}
  \mathcal{F}_{\eta} &=& \int d\mathbf{r} \left [ -\frac{1}{2} \epsilon_\eta n_\eta^2
    + \frac{1}{2} \beta_\eta n_\eta \left ( \nabla^2 + q_\eta^2 \right )^2 n_\eta \right. \nonumber\\
  && \left. - \frac{1}{3} g_\eta n_\eta^3 + \frac{1}{4} v_\eta n_\eta^4 \right ], \label{eq:F_eta}
\end{eqnarray}
while the coupling between $A$ and $B$ components is determined by
\begin{eqnarray}
  \mathcal{F}_{AB} &=& \int d\mathbf{r} \left [ \alpha_{AB} n_A n_B
    + \beta_{AB} n_A \left ( \nabla^2 + q_{AB}^2 \right )^2 n_B \right. \nonumber\\
  && \left. + \frac{1}{2} w_{AB} n_A^2 n_B + \frac{1}{2} u_{AB} n_A n_B^2 \right ]. \label{eq:F_AB}
\end{eqnarray}
All the model parameters ($\epsilon_\eta$, $q_\eta$, $\beta_\eta$, $g_\eta$, $v_\eta$, $\alpha_{AB}$,
$\beta_{AB}$, $q_{AB}$, $w_{AB}$, $u_{AB}$) are dimensionless and can be expressed through
the expansion components of the Fourier transform of two- and three-point direct correlation
functions \cite{Taha19}. In Eq.~(\ref{eq:F_AB}) the first $\alpha_{AB}$ term yields an energy
penalty for the overlap of $A$ and $B$ density maxima (atomic sites), while the last two terms
are important for stabilizing the vacancy positions (i.e., without either $A$ or $B$ sites and with
the overlap of $A$ and $B$ minima). A number of phases of 2D binary ordering plus a homogeneous
state are identified in this model and the corresponding phase diagrams have been calculated
\cite{Taha17,Taha19}, including the coexistence and phase transformation between them. Among
them the binary honeycomb phase, with each of its triangular sublattices occupied by $A$ or
$B$ component separately, corresponds to the lattice structure of monolayer \textit{h}-BN
or the in-plane projection of trigonal prismatic 2H phase of MX$_2$ TMDs, while the phase with
triangular $A(B)$ and honeycomb $B(A)$ sublattices is the in-plane version of octahedral 1T phase
of TMDs.

This PFC model has been used to identify and predict the defect structure, energy, and dynamics
of \textit{h}-BN grain boundaries \cite{Taha17,Waters22}, with results consistent with
experiments and atomistic calculations (DFT or MD), and to examine graphene/\textit{h}-BN
and \textit{h}-BN/\textit{h}-BN heterostructures and bilayers \cite{HirvonenPRB19,ElderPRM21}
as well as thermal transport of \textit{h}-BN monolayers \cite{DongPCCP18}. It is important
to note that the model can also be applied to a wider range of 2D compound materials with
binary honeycomb lattice, including the atomically thin MX$_2$ TMDs of 2H phase
(e.g., M = Mo, W, Nb, X = S, Se; MoTe$_2$, TaS$_2$) and transition metal chalcogenides of
1H phase (e.g., FeSe) \cite{ZhouNature18}. Although in 2H phase a MX$_2$ monolayer is composed
of X-M-X stacking planes, the two X atoms are always paired and thus can be effectively treated
as one base unit occupying a honeycomb lattice together with the M atoms, a configuration that
is well described by this in-plane PFC model particularly for the monolayer deposited on a
substrate during epitaxy (e.g., the out-of-plane corrugations of MoS$_2$ grown on Au(111)
via physical vapor deposition are mostly less than 1 {\AA} \cite{SorensenACSNano14}, and thus
play a secondary role and are neglected in our modeling). This can also be seen from the
result that the defect core structures of grain boundaries (e.g., $4|8$, $4|4$, and $8|8$
dislocations) found in MoS$_2$ \cite{ZouNanoLett13,ZhouNanoLett13,EnyashinJPCC13} and MoSe$_2$
\cite{LiuPRL14} samples can be identified from this 2D PFC model \cite{Taha17}.

Here the above binary model is extended to an in-plane $AB$/$CB$ or $ACB$ ternary PFC
model describing the mixture of $AB$ and $CB$ compounds (each having its own intrinsic $AB$
or $CB$ sublattice ordering) that are confined on a substrate. The corresponding free energy
functional is written as
\begin{equation}
  \mathcal{F} = \mathcal{F}_A + \mathcal{F}_B + \mathcal{F}_C
  + \mathcal{F}_{AB} + \mathcal{F}_{CB} + \mathcal{F}_{AC}, \label{eq:F_ABC}
\end{equation}
where $\mathcal{F}_{\eta=A,B,C}$ is determined by Eq.~(\ref{eq:F_eta}) giving triangular sublattice
for each of $A$, $B$, and $C$ components in the crystalline state, while $\mathcal{F}_{AB}$ and
$\mathcal{F}_{CB}$ follow Eq. (\ref{eq:F_AB}) (with $A \rightarrow C$ for $\mathcal{F}_{CB}$) to
stabilize $AB$ and $CB$ binary honeycomb lattices, respectively. The specific form of
$\mathcal{F}_{CB}$ is written as
\begin{eqnarray}
  \mathcal{F}_{CB} &=& \int d\mathbf{r} \left [ \alpha_{CB} n_C n_B
    + \beta_{CB} n_C \left ( \nabla^2 + q_{CB}^2 \right )^2 n_B \right. \nonumber\\
  && \left. + \frac{1}{2} w_{CB} n_C^2 n_B + \frac{1}{2} u_{CB} n_C n_B^2 \right ]. \label{eq:F_CB}
\end{eqnarray}
The tendency of $A$-$C$ mutual exclusion is built into $\mathcal{F}_{AC}$; to leading order we have
\begin{eqnarray}
  \mathcal{F}_{AC} &=& \int d\mathbf{r} \left [
    \beta_{AC} n_A \left ( \nabla^2 + q_{AC}^2 \right )^2 n_C + \frac{1}{2} \mu_{AC} n_A^2 n_C^2
    \right. \nonumber\\
  && \left. + \frac{1}{2} w_{AC} n_A^2 n_C + \frac{1}{2} u_{AC} n_A n_C^2 \right ],
  \label{eq:F_AC}
\end{eqnarray}
where the parameter $\beta_{AC}$ controls the degree of proximity (or relative affinity) between
$A$ and $C$ species, with larger value of $\beta_{AC}$ ($>0$) corresponding to more energetic
preference of $A$-$C$ heteroelemental bonding or neighboring as compared to $A$-$A$ and $C$-$C$
homoelemental ones. It is noted that the model introduced here is different from the
multicomponent PFC model developed in Ref.~\cite{HirvonenPRB19} for the study of
graphene/\textit{h}-BN phase-separated heterostructure, where the couplings between spatially
smoothed density fields (with the filtering of $n_\eta$ to eliminate short lattice-scale variations)
were imposed to control the phase separation and structure stability. This extra treatment of
density smoothing is not needed here in our modeling of the $AB$/$CB$ compound system. Instead,
a simple high-order coupling term, $\mu_{AC} n_A^2n_C^2$, is introduced in Eq.~(\ref{eq:F_AC})
to favor the separation of $A$ and $C$ atomic sites (with both $n_A$ and $n_C$ maxima) and
also of their vacancy sites (with both $n_A$ and $n_C$ minima). In addition, the term
$\alpha_{AC} n_A n_C$ is neglected here (or equivalently, $\alpha_{AC}=0$). $\alpha_{AC}>0$
corresponds to the energetic favoring of $A$-$C$ heteroelemental bonding, an effect that
has already been incorporated in the $\beta_{AC}$ term, while a more negative value of
$\alpha_{AC}$ would lead to a higher degree of $A$-$C$ site overlap which should be avoided.

This ternary PFC model, including the newly identified $A$-$C$ coupling terms
in Eq.~(\ref{eq:F_AC}), can be derived from DDFT by following the procedure similar to that
given in Ref.~\cite{Taha19} for binary systems. In addition to those two- and three-point
direct correlation functions of classical DFT that are used to identify the PFC terms in
Eqs.~(\ref{eq:F_eta})--(\ref{eq:F_CB}) for $A$-$B$ and $C$-$B$ binary sublattice ordering,
the Fourier-space expansions of two-, three-, and four-point direct correlations between
$A$ and $C$ components are needed to obtain the new $\beta_{AC}$, $w_{AC}$, $u_{AC}$, and
$\mu_{AC}$ terms of $A$-$C$ coupling in Eq.~(\ref{eq:F_AC}), respectively. In principle,
this ternary PFC model can be extended to incorporate out-of-plane deformations of $AB$/$CB$
or $ACB$ monolayers, by using the approach developed in Ref.~\cite{ElderPRM21} with the
coupling to the variation of an additional field of vertical surface height. In the epitaxial
system examined here, with substrate confinement and hence rather weak vertical variations of
the grown overlayers as observed experimentally and described above, out-of-plane deformations
would be of secondary or negligible effect and thus are not considered in this study.

The dynamics of three density variation fields $n_A$, $n_B$, and $n_C$ are governed by
Eq.~(\ref{eq:n_dynamics}). Substituting Eq.~(\ref{eq:F_ABC}), with the use of
Eqs.~(\ref{eq:F_eta}), (\ref{eq:F_AB}), (\ref{eq:F_CB}), and (\ref{eq:F_AC}), gives
\begin{eqnarray}
  \frac{\partial n_A}{\partial t} &=& m_A \nabla^2 \left [ -\epsilon_A n_A
    + \beta_A ( \nabla^2 + q_A^2 )^2 n_A - g_A n_A^2 \right. \nonumber\\
    && + v_A n_A^3 + \alpha_{AB}n_B + \beta_{AB} ( \nabla^2 + q_{AB}^2 )^2 n_B
    \nonumber\\
    && + w_{AB}n_An_B + \frac{1}{2}u_{AB}n_B^2
    + \beta_{AC} ( \nabla^2 + q_{AC}^2 )^2 n_C \nonumber\\
    && \left. + \mu_{AC} n_An_C^2 + w_{AC}n_An_C + \frac{1}{2}u_{AC}n_C^2 \right ], \label{eq:nA}
\end{eqnarray}
\begin{eqnarray}
  \frac{\partial n_B}{\partial t} &=& m_B \nabla^2 \left [ -\epsilon_B n_B
    + \beta_B ( \nabla^2 + q_B^2 )^2 n_B - g_B n_B^2 \right. \nonumber\\
    && + v_B n_B^3 + \alpha_{AB}n_A + \beta_{AB} ( \nabla^2 + q_{AB}^2 )^2 n_A
    \nonumber\\
    && + u_{AB}n_An_B + \frac{1}{2}w_{AB}n_A^2
    + \beta_{CB} ( \nabla^2 + q_{CB}^2 )^2 n_C \nonumber\\
    && \left. + \alpha_{CB}n_C + u_{CB}n_Cn_B + \frac{1}{2}w_{CB}n_C^2 \right ], \label{eq:nB}
\end{eqnarray}
\begin{eqnarray}
  \frac{\partial n_C}{\partial t} &=& m_C \nabla^2 \left [ -\epsilon_C n_C
    + \beta_C ( \nabla^2 + q_C^2 )^2 n_C - g_C n_C^2 \right. \nonumber\\
    && + v_C n_C^3 + \alpha_{CB}n_B + \beta_{CB} ( \nabla^2 + q_{CB}^2 )^2 n_B
    \nonumber\\
    && + w_{CB}n_Cn_B + \frac{1}{2}u_{CB}n_B^2
    + \beta_{AC} ( \nabla^2 + q_{AC}^2 )^2 n_A \nonumber\\
    && \left. + \mu_{AC} n_A^2n_C + u_{AC}n_An_C + \frac{1}{2}w_{AC}n_A^2 \right ], \label{eq:nC}
\end{eqnarray}
which are used in our simulations of 2D ternary $AB$/$CB$ type material systems.

\begin{figure*}
  \centerline{\includegraphics[width=\textwidth]{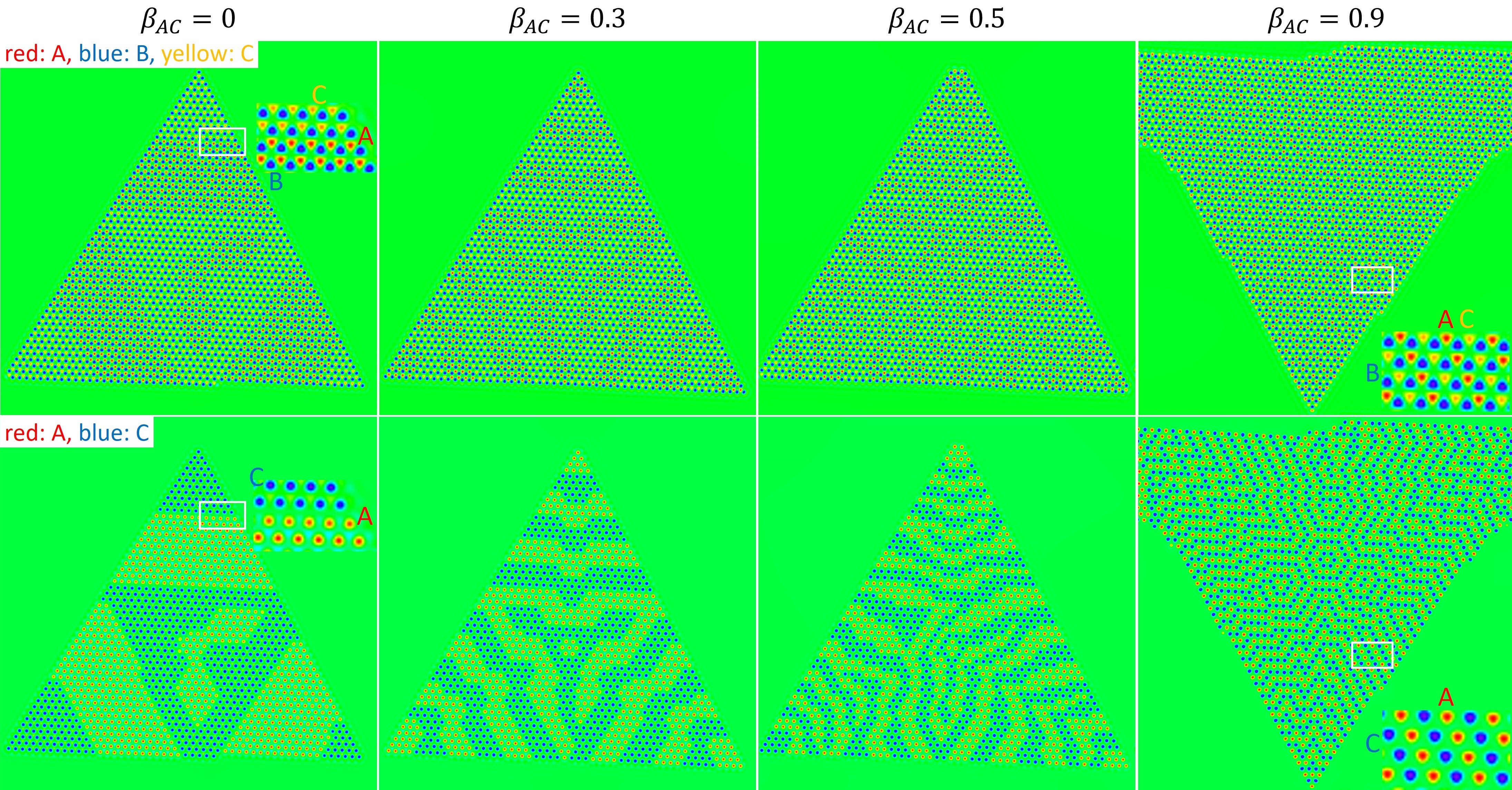}}
  \caption{Sample simulation snapshots of ternary 2D single crystals with hexagonal lattice symmetry
    surrounded by a homogeneous state, showing a transition from phase separation to atomic
    ordering between $A$ and $C$ species as the value of $\beta_{AC}$ increases. Results are
    generated at $t=10^4$ for $\epsilon=0.3$ and average densities $n_{A0}=n_{B0}=n_{C0}=-0.383$,
    starting from an initial single nucleus with only one $A$ and one $B$ atoms embedded in the
    homogeneous media. Top panels: Spatial distributions of all three densities $n_A$, $n_B$,
    and $n_C$. Bottom panels: The corresponding spatial distributions of $n_A$ and $n_C$ only
    (noting the change of $C$-site coloring from yellow to blue as compared to the upper
    panels for a better contrast). Some portions of the simulated atomic configurations
    (white-boxed) are enlarged as insets, showing either phase-separated or
    atomically ordered lattice structures.}
  \label{fig:phasesep-ordering}
\end{figure*}

\section{Results}

The PFC dynamical equations (\ref{eq:nA})--(\ref{eq:nC}) are solved numerically via a
pseudospectral method with the imposing of periodic boundary conditions, starting from 
various initial conditions. For simplicity, the in-plane systems studied in this work are
free of heterointerfacial defects or morphological modulations, as found in many $AB$/$CB$
type experimental systems (e.g., TMD/TMD heterostructures) with small enough lattice mismatch
so that the misfit-induced effects can be neglected at least to lowest order. This will
enable us to focus here on the properties of phase separation or demixing, atomic ordering,
intermixing, and the formation of lateral heterostructures and multijunctions through
PFC modeling. Results for more complex scenarios caused by misfit strains, such as the
stress-driven 
structural variations or dislocation formation which can also be well addressed by this
PFC model, will be presented elsewhere. In the following the model parameters are set as
$\beta_\eta=1$, $q_\eta=1$, $g_\eta=0.5$, $v_\eta=1$, and $m_\eta=1$ (with $\eta=A,B,C$),
as well as $\alpha_{AB}=\alpha_{CB}=0.5$, $\beta_{AB}=\beta_{CB}=0.02$, $q_{AB}=q_{CB}=q_{AC}=1$,
$\mu_{AC}=1$, and $w_{AB}=w_{CB}=w_{AC}=u_{AB}=u_{CB}=u_{AC}=0.3$. Values of parameter $\beta_{AC}$
characterizing $A$-$C$ heteroelemental interaction, the effective temperature parameter
$\epsilon_A=\epsilon_B=\epsilon_C=\epsilon$, and average densities $n_{\eta0}$ are
varied to represent different growth and sample conditions.

\subsection{Effect of heteroelemental interaction: Atomic ordering versus phase separation}
\label{sec:order_sep}

Analogous to the well-known scenarios of phase separation (demixing) vs short-range lattice
ordering in binary alloying systems, it is expected that the 2D $AB$/$CB$ ternary materials
studied here, with sublattice-ordered structure in each of the $AB$ and $CB$ compounds, would
reveal a similar behavior giving either the separation between $AB$ and $CB$ honeycomb phases
or an atomically ordered phase with an additional lattice ordering between $A$ and $C$ components.
This effect has been built into our PFC model via the $\beta_{AC}$ term in Eq.~(\ref{eq:F_AC})
representing the degree of energetic preference of $A$-$C$ heteroelemental bonding over $A$-$A$
and $C$-$C$ homoatomic interactions, and is shown explicitly in our simulation results
illustrated in Fig.~\ref{fig:phasesep-ordering}. Here the nucleated growth of 2D ternary
crystallites is simulated, as initiated from a single nucleus and surrounded by a homogeneous
phase of $A$, $B$, and $C$ densities with $n_{A0}=n_{B0}=n_{C0}=-0.383$ at $\epsilon=0.3$.
Inside a 2D crystal sheet of triangle shape (as found in most
experiments of TMD monolayers grown epitaxially), the phase-separated state with unmixing $AB$
and $CB$ domains occurs at small enough $\beta_{AC}$, while the average size or length scale
of the stripe-like domain pattern reduces with the increase of $\beta_{AC}$. At large enough
value of $\beta_{AC}$, i.e., large enough energetic preference for heteroatomic coordination,
a transition to the $A$-$C$ atomic ordering occurs, showing as a fully ordered lattice structure
with $A$ and $C$ components forming a triangular lattice while each of $AB$ and $CB$ still
maintaining its binary honeycomb atomic configuration [see both the $A$-$B$-$C$ (upper panel)
and $A$-$C$ (bottom panel) density distributions in the insets of Fig.~\ref{fig:phasesep-ordering}
for $\beta_{AC}=0.9$]. This is of the same atomically ordered structure observed in the
recent experiment of 2D Re$_{0.5}$Nb$_{0.5}$S$_2$ TMD alloy, resembling a geometrically frustrated
system with Re($A$)-Nb($C$) triangular lattice \cite{AziziPRL20}.

\begin{figure}
  \centerline{\includegraphics[trim=0 40 60 70,clip,width=0.5\textwidth]{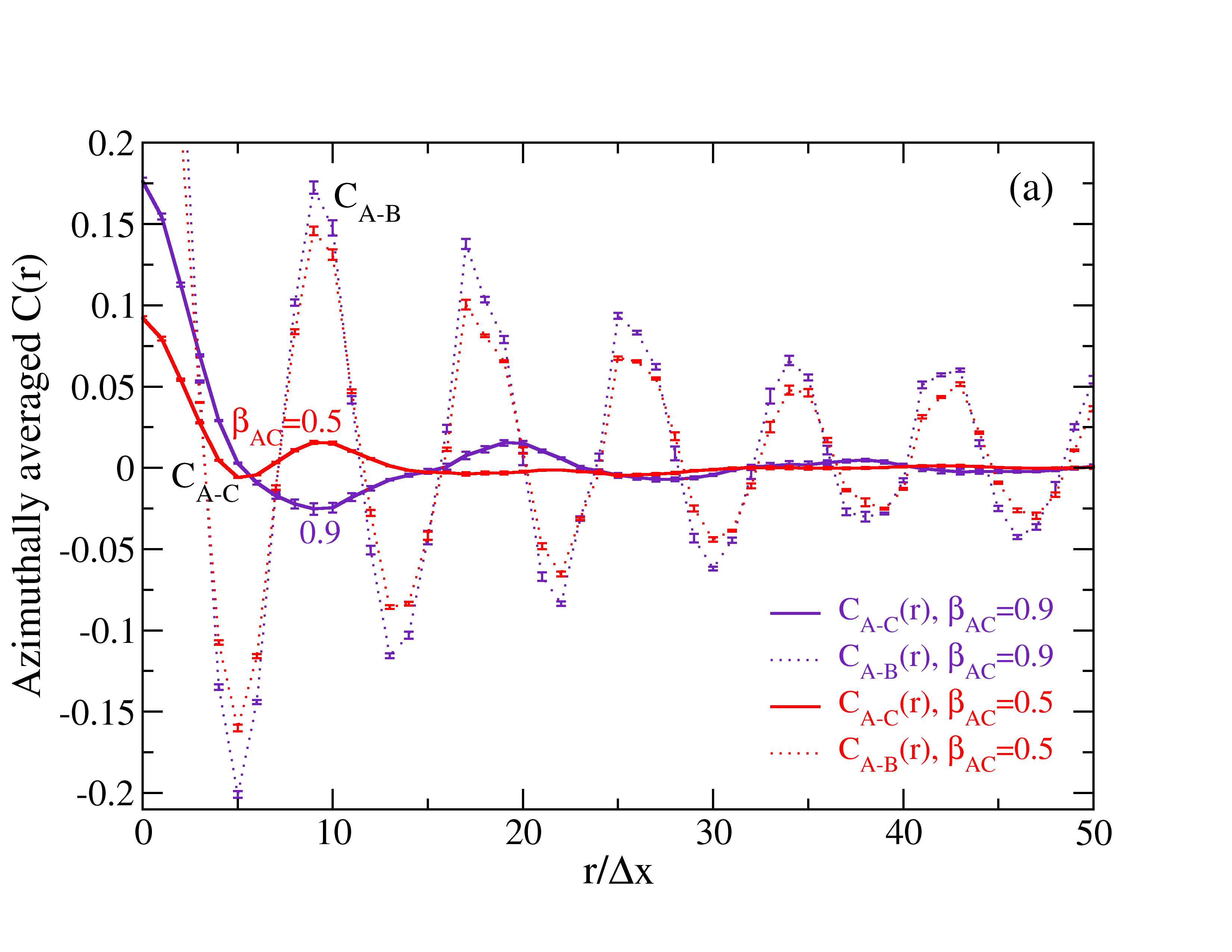}}
  \centerline{\includegraphics[trim=0 40 60 70,clip,width=0.5\textwidth]{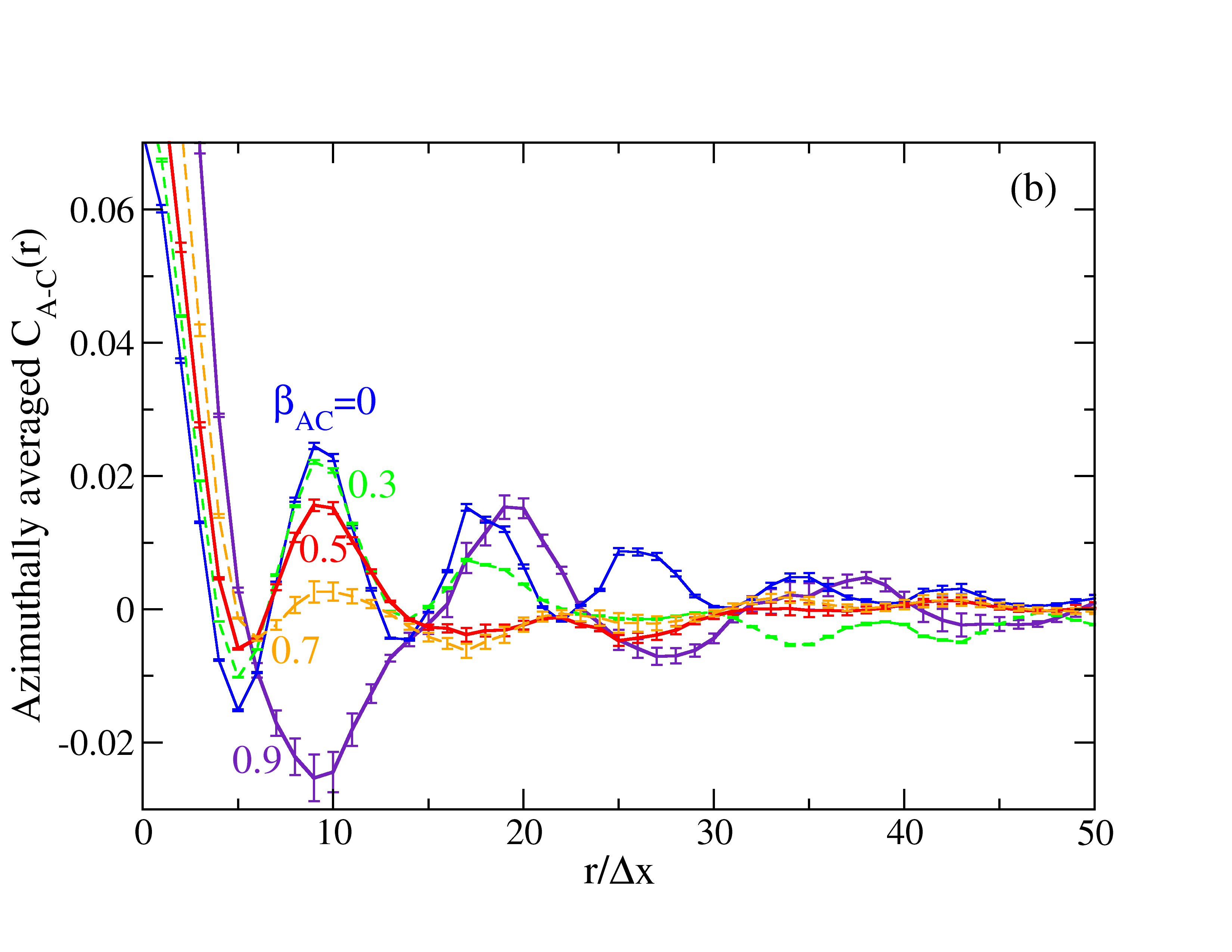}}
  \caption{Azimuthally averaged correlation functions for different values of $\beta_{AC}$
    at $\epsilon=0.3$ and $n_{A0}=n_{B0}=n_{C0}=-0.383$. (a) Correlation functions $C_\text{A-C}(r)$ 
    and $C_\text{A-B}(r)$ for $\beta_{AC}=0.5$ and $0.9$. (b) Results of $C_\text{A-C}(r)$ 
    for $\beta_{AC}=0$, $0.3$, $0.5$, $0.7$, and $0.9$ [note the much smaller scale of the
    vertical axis as compared to (a)]. Some of the corresponding atomic structures are given
    in Fig.~\ref{fig:phasesep-ordering}. The data has been averaged over 20 simulations per
    each value of $\beta_{AC}$.}
  \label{fig:corr_betaAC}
\end{figure}

The property of lattice ordering can be quantified via the equal-time pair correlation function
\begin{eqnarray}
  && C_{\eta\text{-}\eta'}(\mathbf{r},t) \nonumber\\
  && = \left \langle \left [ n_\eta(\mathbf{x}+\mathbf{r},t) - n_{\eta'}(\mathbf{x}+\mathbf{r},t) \right ]
  \left [ n_\eta(\mathbf{x},t) - n_{\eta'}(\mathbf{x},t) \right ] \right \rangle \nonumber\\
  && \quad - \langle n_\eta - n_{\eta'} \rangle^2, \label{eq:corr}
\end{eqnarray}
where $\eta \neq \eta'$ and $\eta, \eta' = A, B, C$. In the crystalline state the density fields
$n_\eta$ and $n_{\eta'}$ vary periodically in space, and a positive maximum of spatial correlation
$C_{\eta\text{-}\eta'}$ at a displacement $\mathbf{r}$ indicates a homoelemental $\eta$-$\eta$ or
$\eta'$-$\eta'$ pair of atomic sites with separation of $\mathbf{r}$, while a negative minimum
of $C_{\eta\text{-}\eta'}$ corresponds to a heteroelemental $\eta$-$\eta'$ atomic pair instead.
Some results of the azimuthal average of $C_{\eta\text{-}\eta'}$, including $C_\text{A-C}(r)$ and
$C_\text{A-B}(r)$ at time $t=10^4$ (each averaged over 20 independent simulation runs
initialized with different random number seeds for the homogeneous media),
are given in Fig.~\ref{fig:corr_betaAC}. As expected, the first negative minimum of the
oscillatory $C_\text{A-B}(r)$ shown in Fig.~\ref{fig:corr_betaAC}(a) is located at
$r=a_{AB} \sim 5\Delta x$ (with the simulation grid spacing $\Delta x = \pi/4$) which is the
distance of $A$-$B$ nearest neighboring in an $AB$ honeycomb unit ring, and the next positive
maximum of $C_\text{A-B}(r)$ appears at $r=a_0 \simeq \sqrt{3}a_{AB}$ corresponding to either
$A$-$A$ or $B$-$B$ neighboring, i.e., the lattice spacing of binary $AB$ honeycomb structure.
Similar results for the $CB$ honeycomb lattice ordering can be obtained from the correlation
$C_\text{C-B}$.

Quantitative information for the $A$-$C$ segregation or ordering can be extracted from the
correlation function $C_\text{A-C}(r)$. As seen in Fig.~\ref{fig:corr_betaAC}(b), large enough
value of $\beta_{AC}$ (e.g., $=0.9$) leads to a negative minimum of $C_\text{A-C}$ at $r=a_0$,
which corresponds to the nearest-neighbor spacing of $A$-$C$-$A$ or $C$-$A$-$C$ triangular
lattice with more $A$-$C$ heteroatomic neighboring that results in the negative value of spatial
correlation. This agrees with the result of Re-Nb spatial correlation in Re$_{0.5}$Nb$_{0.5}$S$_2$
monolayer measured in Ref.~\cite{AziziPRL20}. The next positive maximum of correlation is
found at $r \simeq 2a_0$, which can be attributed to the homoatomic next-nearest-neighbor
($A$-$A$ or $C$-$C$) of the binary $AC$ lattice. These then indicate the $A$-$C$ atomic
ordering with $ACAC...$ alternative lines of atoms as shown in the bottom-right inset of
Fig.~\ref{fig:phasesep-ordering}. On the other hand, Fig.~\ref{fig:corr_betaAC}(b) also
shows that at smaller $\beta_{AC}$ (e.g., $0 \leq \beta_{AC} \leq 0.7$) a positive maximum
of $C_\text{A-C}$ instead occurs at $r=a_0$, indicating the dominance of homoelemental
$A$-$A$ or $C$-$C$ lattice inside each phase-segregated domain. The height of this
maximum peak decreases with the increase of $\beta_{AC}$, due to the contribution of negative
correlation from larger portion of $A$-$C$ heterointerfaces separating domains of smaller size.
This domain size reduction is corroborated by the less number of correlation peaks located at
larger distances $r$ and also shorter range of positive spatial correlation (or faster decay
of the envelop of positive correlation peaks) when $\beta_{AC}$ becomes larger. All these
results are consistent with the simulation outcomes presented in Fig.~\ref{fig:phasesep-ordering}
for a transition between states of phase separation and short-range atomic ordering.

\subsection{Intermixing and disordering}
\label{sec:intermixing}

\begin{figure}
  \centerline{\includegraphics[clip,width=0.45\textwidth]{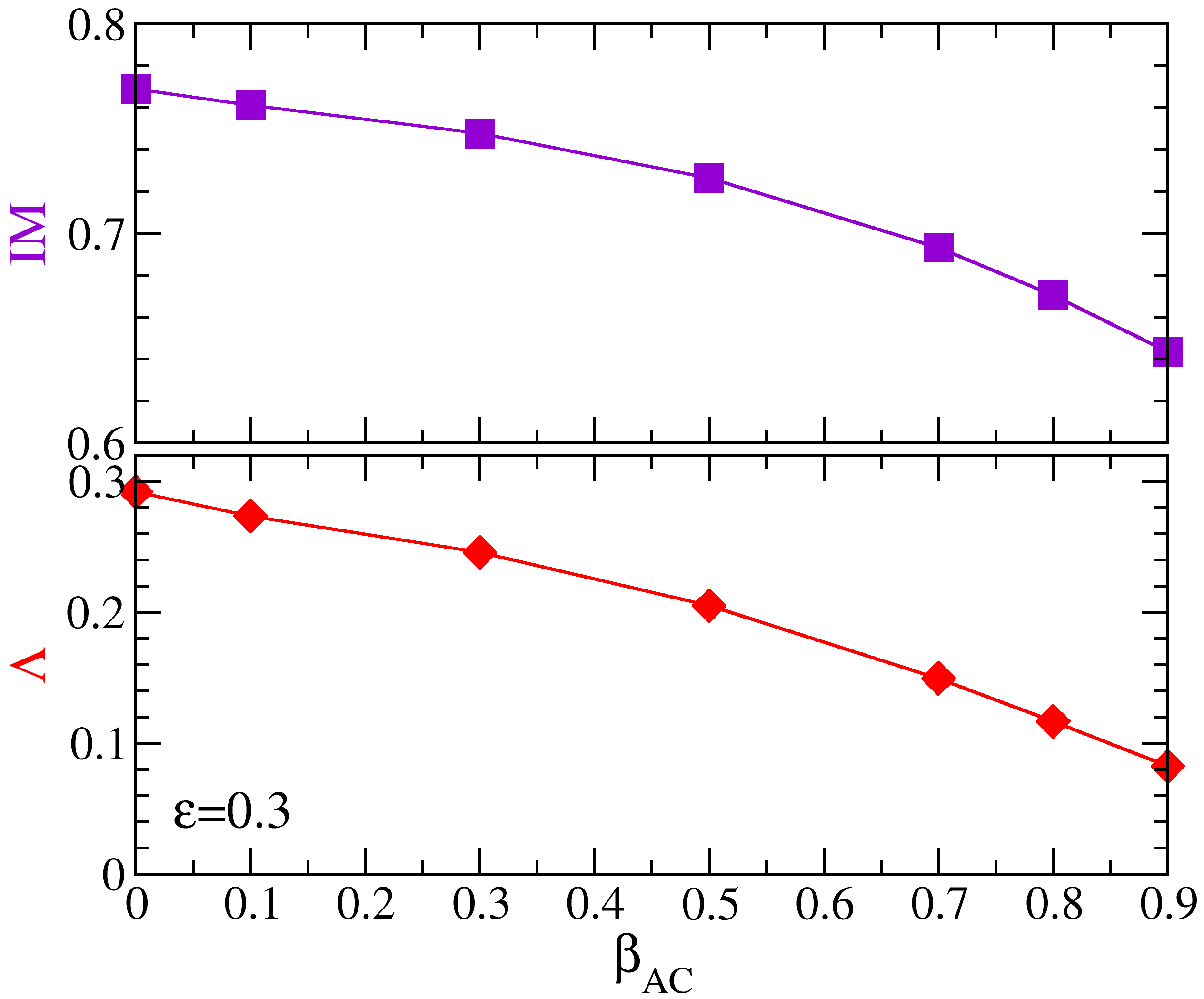}}
  \caption{(a) The intermixing factor IM and (b) the normalized overlap factor $\Lambda$
    as a function of $\beta_{AC}$ at $\epsilon=0.3$. The results have been averaged over 20
    simulations (with error bars smaller than the size of symbols shown). Some of the corresponding 
    atomic structures and correlation functions are given in Figs.~\ref{fig:phasesep-ordering}
    and \ref{fig:corr_betaAC}, respectively.}
  \label{fig:IM_betaAC}
\end{figure}

During the growth and evolution of 2D ternary TMD monolayers, the compositional intermixing or
disordering of $AB$ vs $CB$ compound has been observed, showing as the random distribution or
alloying of $A$ or $C$ components in the experimental samples (e.g., Mo$_{1-x}$W$_x$S$_2$
\cite{ChenACSNano13,BogaertNanoLett16}). In our PFC modeling this behavior is represented by
the degree of intermixing between $A$ and $C$ species (or the probability of $A$-$C$ intermixing).
This can be identified quantitatively via the following two factors. The first one is an intermixing
factor (IM) at a given time $t$, which we define as
\begin{equation}
  \text{IM} = 1 - \frac{\left \langle \left [ \delta n_A(\mathbf{r}) - \delta n_C(\mathbf{r}) \right ]^2
    \right \rangle} {2 \left [ \langle \delta n_A^2(\mathbf{r}) \rangle + \langle \delta n_C^2(\mathbf{r})
      \rangle \right ]}, \label{eq:IM}
\end{equation}
where $\delta n_{A(C)} = n_{A(C)} - \langle n_{A(C)} \rangle$ and $\langle ... \rangle$ corresponds
to the spatial average over position $\mathbf{r}$. In the case of complete intermixing or
density overlap (with equal probability of $A$ and $C$ species occupying the same position),
$\delta n_A = \delta n_C$ and thus $\text{IM}=1$. In the other limit of no density overlap,
i.e., $\int \delta n_A(\mathbf{r}) \delta n_C(\mathbf{r}) d\mathbf{r} = 0$ and hence
$\langle \delta n_A(\mathbf{r}) \delta n_C(\mathbf{r}) \rangle = 0$, we have $\text{IM}=1/2$
without any intermixing. (Note that $\text{IM}=0$ corresponds to the inverse atomic ordering
between $A$ and $C$ with $\delta n_A = -\delta n_C$, such as the triangular $A$ (or $C$) and
honeycomb or inverse triangular $C$ (or $A$) sublattice ordering \cite{Taha19}, which can be
viewed as the in-plane projection of metallic 1T phase of TMDs and is not studied here.)

\begin{figure}
  \centerline{\includegraphics[clip,width=0.5\textwidth]{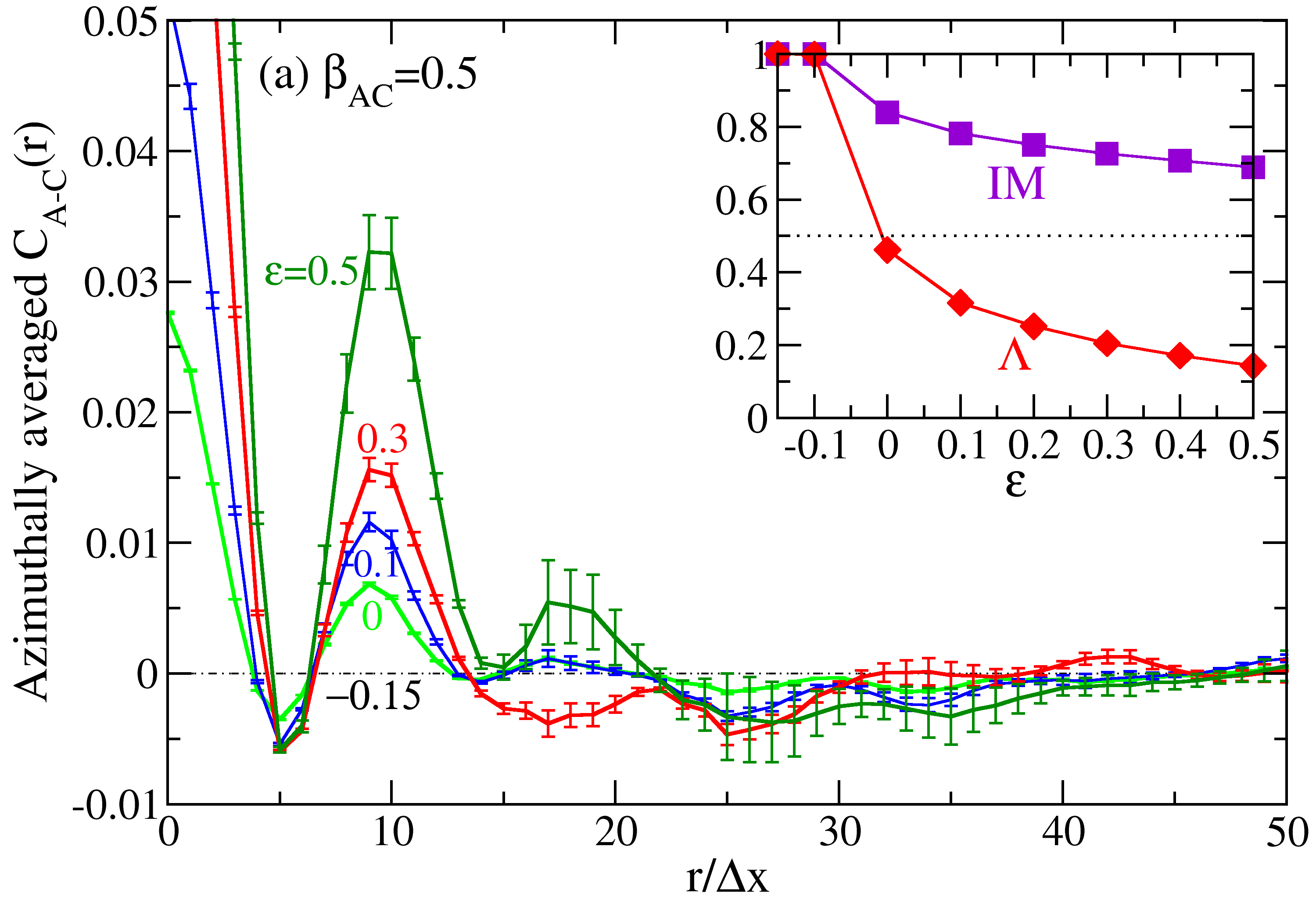}}
  \centerline{\includegraphics[clip,width=0.5\textwidth]{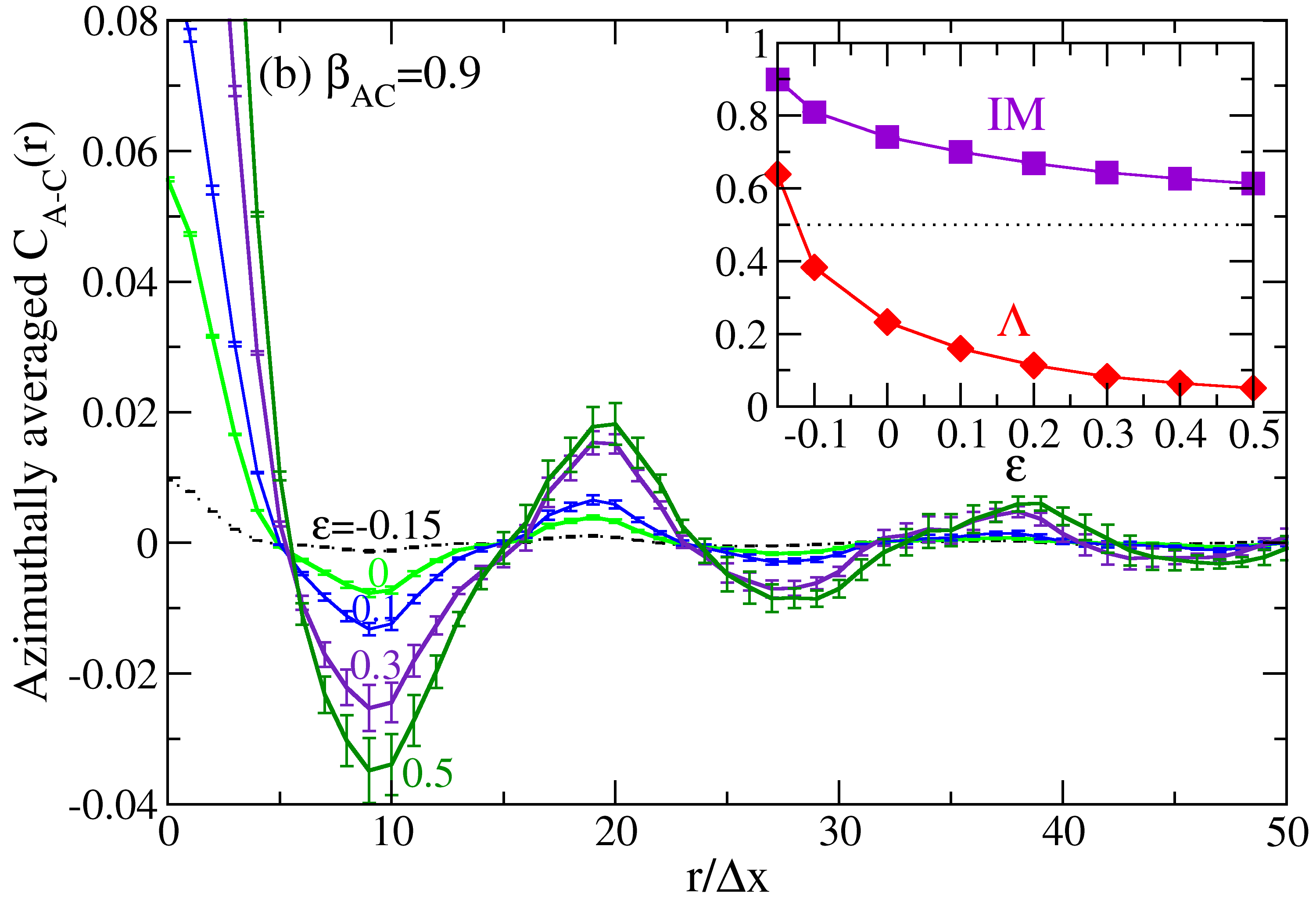}}
  \caption{Azimuthally averaged correlation function $C_\text{A-C}(r)$ for different values
    of $\epsilon$ at (a) $\beta_{AC}=0.5$ and (b) $\beta_{AC}=0.9$. The corresponding results
    for the degree of $A$-$C$ intermixing are shown in the insets (where the error bars are
    smaller than the size of symbols). All the results have been averaged over 20 simulations.}
  \label{fig:corr_IM_eps}
\end{figure}

Alternatively, an overlap factor $\Lambda$ can be also used to quantify the intermixing, i.e.,
\begin{equation}
  \Lambda = \frac{\left \langle \delta n_A(\mathbf{r}) \delta n_C(\mathbf{r}) \right \rangle^2}
          {\langle \delta n_A^2(\mathbf{r}) \rangle \langle \delta n_C^2(\mathbf{r}) \rangle},
\end{equation}
which is similar to the normalized overlap integral used in the study of mixing or demixing of
binary species \cite{JainPRA11}. A full degree of intermixing (with complete density overlap
$\delta n_A = \delta n_C$) leads to $\Lambda=1$, while the complete lack of intermixing
(with no overlap $\int \delta n_A(\mathbf{r}) \delta n_C(\mathbf{r}) d\mathbf{r} = 0$) yields
$\Lambda=0$. Generally this overlap factor $\Lambda$ might give a better
resolution for quantifying the degree of intermixing as compared to the intermixing factor
IM described above (see, e.g., our calculation results in Figs.~\ref{fig:IM_betaAC} and
\ref{fig:corr_IM_eps}), due to a broader range of $0 \leq \Lambda \leq 1$ as compared to
$1/2 \leq \text{IM} \leq 1$. However, it should be noted that an ambiguity would occur when
using $\Lambda$ in the case of inverse sublattice ordering ($\delta n_A = -\delta n_C$) which
also leads to $\Lambda=1$. This ambiguity can be clarified through the combination with
the IM calculation (which yields $\text{IM}=0$ for complete inverse ordering), and thus
both $\Lambda$ and IM are used in our quantitative analyses of intermixing.

Some results of IM and $\Lambda$ for various values of $\beta_{AC}$ at $\epsilon=0.3$
(which correspond to the simulations conducted in Sec.~\ref{sec:order_sep} and
Figs.~\ref{fig:phasesep-ordering} and \ref{fig:corr_betaAC}) are presented in
Fig.~\ref{fig:IM_betaAC}, showing a small or moderate degree of intermixing. With the
increase of $\beta_{AC}$ (i.e., more energetic favoring of $A$-$C$ heteroelemental coordination)
as accompanied by the transition from $AB$-$CB$ phase separation to atomic ordering, both
values of IM and $\Lambda$ decrease, indicating a lesser intermixing during the transition.

Importantly, our modeling also reveals an increased degree of intermixing or disordering
(random alloying) between $A$ and $C$ species at higher growth temperature, consistent with
that observed in recent experiments of TMD growth \cite{BogaertNanoLett16,ChiuAdvMater18}.
This is shown in the insets of Fig.~\ref{fig:corr_IM_eps}, where values of IM and $\Lambda$
become larger with the decrease of parameter $\epsilon$, i.e., the increase of temperature.
It can also be seen from the azimuthally averaged correlation function $C_\text{A-C}(r)$ plotted
in Figs.~\ref{fig:corr_IM_eps}(a) and \ref{fig:corr_IM_eps}(b), for locally phase-segregated
(with $\beta_{AC}=0.5$) and atomically ordered (with $\beta_{AC}=0.9$) cases respectively.
The height of the first correlation maximum (peak) or minimum (valley) is reduced via lowering
the value of $\epsilon$, indicating less degree of $A$-$C$ correlation at spacing $r=a_0$ of
the $A$ and/or $C$ lattice and hence more disordering of the two species as a result of their
mixing at higher temperature. The connection between $A$-$C$ correlation and intermixing can
be also obtained by rewriting Eq.~(\ref{eq:corr}) at a given time $t$ as
\begin{eqnarray}
  && C_\text{A-C}(\mathbf{r}) \nonumber\\
  && = \left \langle \left [ \delta n_A(\mathbf{x}+\mathbf{r}) -
    \delta n_C(\mathbf{x}+\mathbf{r}) \right ] \left [ \delta n_A(\mathbf{x}) -
    \delta n_C(\mathbf{x}) \right ] \right \rangle, ~\quad \label{eq:corr_A-C}
\end{eqnarray}
such that $\text{IM} = 1-C_\text{A-C}(0)/2(\langle\delta n_A^2 \rangle + \langle\delta n_C^2\rangle)$
from Eq.~(\ref{eq:IM}). Thus, a higher degree of intermixing at smaller $\epsilon$ (higher
temperature) would lead to smaller $C_\text{A-C}(0)$, as verified in Fig.~\ref{fig:corr_IM_eps}.
In the limit of full intermixing with $\delta n_A \rightarrow \delta n_C$, $A$ and $C$ densities
are then uncorrelated, i.e., $C_\text{A-C}(r) \rightarrow 0$ according to Eq.~(\ref{eq:corr_A-C}),
as seen from the plots of $\epsilon=-0.15$ in Fig.~\ref{fig:corr_IM_eps}.

\begin{figure}
  \centerline{\includegraphics[clip,width=0.5\textwidth]{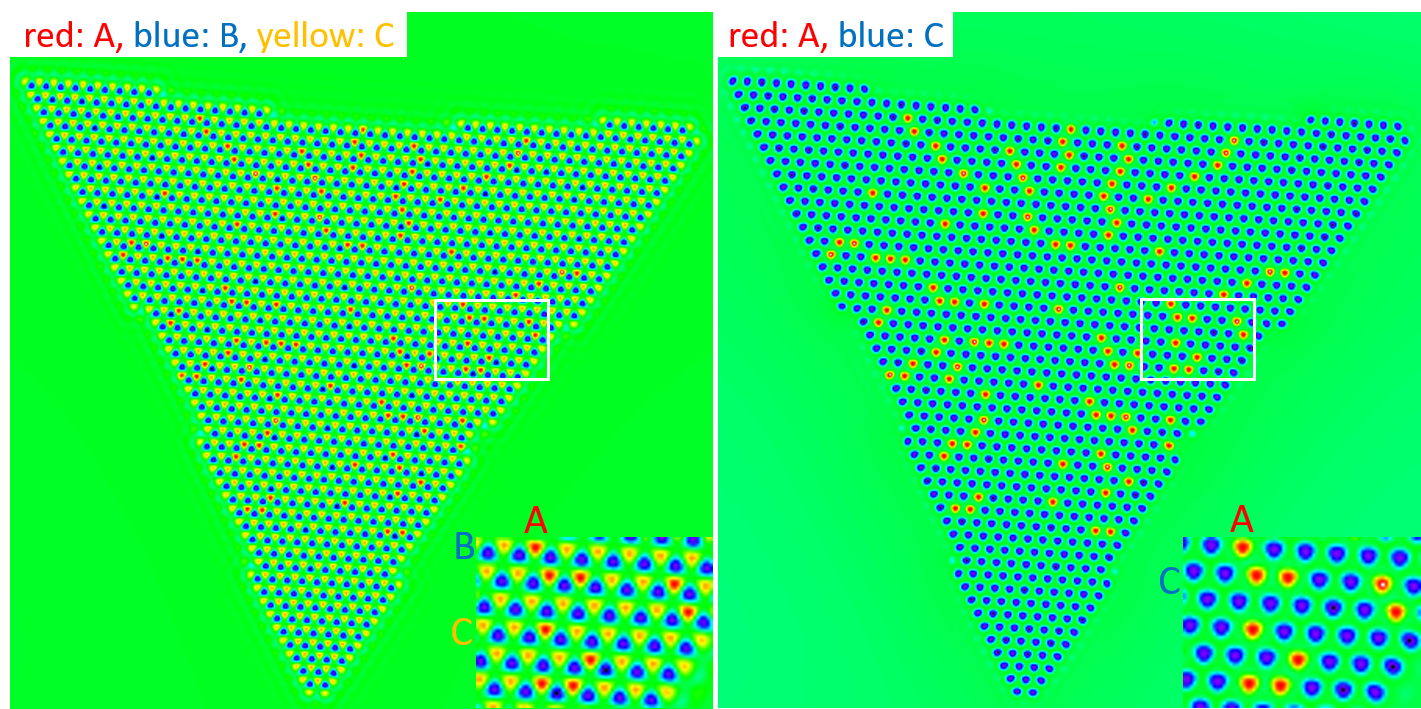}}
  \caption{The spatial distribution of $A$, $B$, and $C$ species (left) and the corresponding
    $A$-$C$ lattice structure (right) obtained from PFC simulation at $\epsilon=0.3$,
    $\beta_{AC}=0.9$, and $\psi_0=-0.6$ with $n_{A0}=n_{B0}=-0.4532$ and $n_{C0}=-0.3128$.
    The white-boxed regions are enlarged as insets.}
  \label{fig:disordered}
\end{figure}

All the above results are for equal composition of $A$ and $C$ components with $n_{A0}=n_{C0}$.
It is expected that at large enough composition disparity between the two species, a dispersed
or random distribution of the minority species would occur. This binary disordered phase can
be reproduced from our PFC simulations as well, with some sample results given in
Fig.~\ref{fig:disordered}. Here we define the concentration of $A$ component via
$c_A = (1+\psi_0)/2$ with
\begin{equation}
  \psi_0 = \frac{\bar{\rho}_A-\bar{\rho}_C}{\bar{\rho}_A+\bar{\rho}_C}
  = \frac{n_{A0}-n_{C0}}{1+n_{A0}+n_{C0}},
\end{equation}
where $\bar{\rho}_{A(C)}$ is the average of atomic number density $\rho_{A(C)}$ of $A$ or $C$
component. This is based on the definition of density variation fields
$n_{A(C)}=(\rho_{A(C)}-\rho_{A(C)0})/\rho_0$ with $\rho_{A(C)0}$ the reference-state densities
and $\rho_0=\rho_{A0}+\rho_{C0}$, the choice of same reference state $\rho_{A0}=\rho_{C0}$,
and $n_{A(C)0} = (\bar{\rho}_{A(C)}-\rho_{A(C)0})/\rho_0$. For the example of $\psi_0=-0.6$
(with $c_A=0.2$) at $\epsilon=0.3$ and $\beta_{AC}=0.9$, Fig.~\ref{fig:disordered} shows
an overall disordered $A$-$C$ structure, while $AB$ or $CB$ still maintains its own binary
honeycomb lattice (see the inset in the left panel of Fig.~\ref{fig:disordered}).

\subsection{Heterostructures and multijunctions via lateral edge-epitaxy}
\label{sec:heterostruc}

\begin{figure*}
  \centerline{\includegraphics[width=\textwidth]{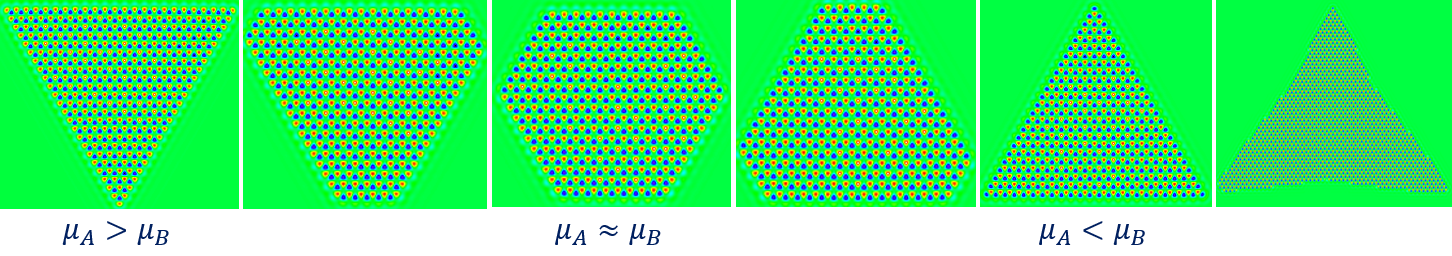}}
  \caption{Grain shape control in PFC simulations of $AB$ binary honeycomb lattice, from
    triangle, truncated triangle, to hexagon shape, through the variation of chemical
    potentials $\mu_A$ and $\mu_B$ (via varying $n_{A0}$ and $n_{B0}$) at
    $\epsilon_A=\epsilon_B=0.3$. All the grains are of zigzag edges, with the density
    maxima of $A$ and $B$ components shown in red and blue, respectively.}
  \label{fig:AB_grains}
\end{figure*}

A starting point of our modeling of 2D heterostructural growth is the understanding of
individual grain growth dynamics, based on some basic mechanisms and outcomes revealed in
the study of binary $AB$ grains \cite{Taha17,Waters22}. As shown in Fig.~\ref{fig:AB_grains},
the grain shape can be controlled via chemical potentials $\mu_A$ and $\mu_B$ of $A$ and $B$
components in the PFC modeling, ranging from triangle, truncated triangle, to hexagon, and
to more irregular shape with faceted surface consisting of terraces. The grain edges are
along the zigzag direction of the honeycomb lattice as obtained from our simulations.
When the conserved dynamics for density fields $n_A$ and $n_B$ are used
and most of $A$-$B$ model parameters remain unchanged, the variation of
$\mu_A$ ($=\delta \mathcal{F} / \delta n_A$) and $\mu_B$ ($=\delta \mathcal{F} / \delta n_B$)
can be effectively tuned by changing the values of average densities $n_{A0}$ and $n_{B0}$,
with $n_{A0} > n_{B0}$ (or $n_{A0} < n_{B0}$) corresponding to $\mu_A > \mu_B$
(or $\mu_A < \mu_B$), as confirmed in numerical simulations.
In addition to grain shape control through the nucleated growth from an initial solid seed
(leading to the results presented in Fig.~\ref{fig:AB_grains}), we are also able to change
the shape of any as-grown grain via varying the relative average densities (i.e., relative
chemical potentials) of the two components, so that the subsequent growth of new $AB$ layers
will transform the binary grain from its initial shape to a different one governed by the
imposed relation of chemical potentials, while the $AB$ sublattice ordering of the grain
microstructure still maintains. Our simulation results are consistent with experiments
\cite{WangChemMater14,SorensenACSNano14,ZhangAdvMater19} and first-principles DFT
calculations \cite{SchweigerJCatal02} of 2D binary TMD materials, and have built the ground
for the subsequent growth of lateral heterostructures, as detailed below.

\begin{figure*}
  \centerline{\includegraphics[width=\textwidth]{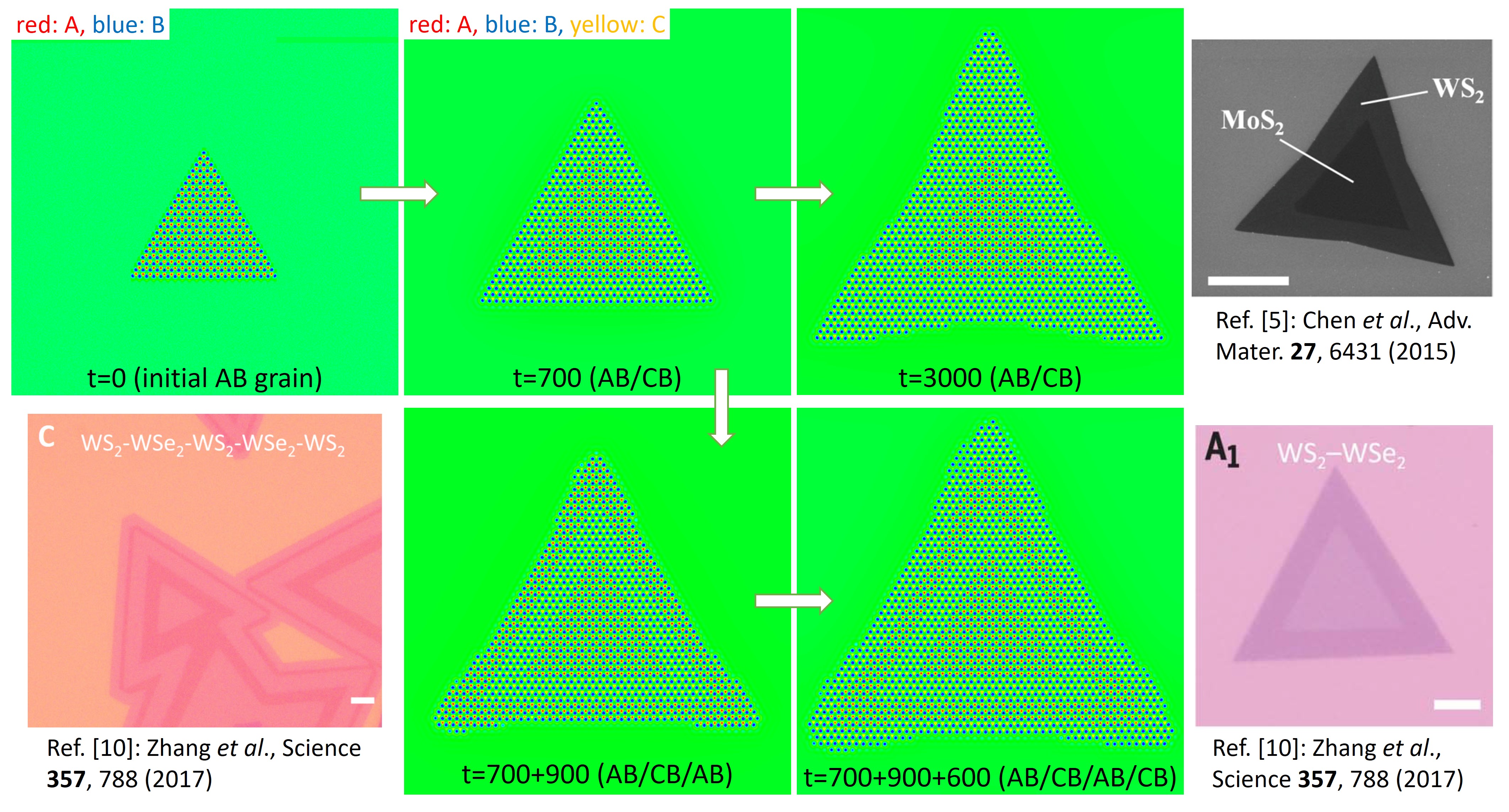}}
  \caption{PFC modeling for the growth of $AB/CB$ lateral heterostructures and $AB/CB/AB$ and
    $AB/CB/AB/CB$ multijunctions or superlattices at $\epsilon=0.3$ and $\beta_{AC}=0.9$,
    with comparison to some recent experimental results (reprinted with permission from
    Ref.~\cite{ChenAdvMater15} Copyright 2015 Wiley-VCH and from Ref.~\cite{ZhangScience17}
    Copyright 2017 AAAS).}
  \label{fig:hetero_ABCB}
\end{figure*}

Here we emulate the experimental growth process of lateral edge-epitaxy, in which the specific
configurations of heterostructures with domain composition segregation are grown sequentially
via flux control. Some sample simulation results and the comparison to some experiments (e.g.,
Refs.~\cite{ChenAdvMater15,ZhangScience17}) are shown in Fig.~\ref{fig:hetero_ABCB}.
The model parameters are chosen to represent typical growth conditions, including $\epsilon=0.3$
for low enough temperature to avoid substantial compositional intermixing across the heterointerface,
and $\beta_{AC}=0.9$ giving an intrinsic trend of atomic ordering (but not phase separation)
in the bulk state of 2D ternary crystal as found in various ternary TMD monolayers
\cite{GanSciRep14,YangChemMater18,AziziPRL20}. The initial condition is a pre-grown triangle-shaped
$AB$ grain (see, e.g., the first panel of Fig.~\ref{fig:hetero_ABCB} at $t=0$), as prepared
according to the binary grain growth mechanism illustrated in Fig.~\ref{fig:AB_grains}. In our PFC
modeling, to facilitate the subsequent growth of $CB$ compounds instead of $AB$ we also initialize
a homogeneous state of $n_C$ with large enough average density $n_{C0}$ throughout the system, while
outside the $AB$ crystalline grain setting an initial homogeneous $n_A$ phase with small enough
$n_{A0}$ and thus low enough density of $A$-type precursors to prevent the formation of unwanted $AB$
layers or interface alloying or intermixing. Specifically, to generate results given in the upper
panels of Fig.~\ref{fig:hetero_ABCB} we first set $n_{A0}=-0.55$ and $n_{B0}=-0.375$ outside the
initial $AB$ grain, and $n_{C0}=-0.375$. The subsequent simulation shows the $CB$ layer (of binary
honeycomb lattice structure) grows epitaxially and laterally from the zigzag $AB$ edge, forming an
in-plane heterostructural grain with sharp and defect-free $AB/CB$ heterointerface as well as
faceted outer surface, consistent with the experimental findings of TMD lateral heterostructures
in 2D triangle-shape crystals \cite{HuangNatMater14,GongNatMater14,DuanNatNanotech14,LiScience15,
ChenAdvMater15,BogaertNanoLett16,ZhangAdvMater18,ZhangAdvMater19}.

The subsequent growth of $AB/CB/AB$ type multijunctions requires a switch from the $C$-component
rich to $A$-component dominated deposition flux, which can be effectively implemented by increasing
the $A$-species density while reducing the $C$-species one (e.g., changing to $n_{A0}=n_{B0}=-0.375$
and $n_{C0}=-0.58$) outside the previously grown $AB/CB$ crystallite. A step-by-step lateral epitaxial
growth of the new $AB$ layer from the outer zigzag edge of the 2D crystal then occurs, leading to
the formation of planar multiple quantum wells (see the bottom panel of Fig.~\ref{fig:hetero_ABCB}).
A similar procedure can be followed to sequentially grow the in-plane superlattice consisting of
alternative types of $AB$ or $CB$ blocks (each still forming its own binary honeycomb lattice).
For the example of $AB/CB/AB/CB$ superlattice shown in Fig.~\ref{fig:hetero_ABCB}, a depleted
density of $A$ component (with $n_{A0}=-0.59$ to avoid the nucleation of any new $AB$ lattice)
and high enough densities for $C$ and $B$ (with $n_{C0}=n_{B0}=-0.375$) are set in the initial
homogeneous media outside the as-grown $AB/CB/AB$ grain, followed by the edge-epitaxy of the
new, outer $CB$ layer. All these simulation outcomes well agree with the experimentally
observed single-crystalline monolayers of coherently modulated lateral multi-heterostructures
or superlattices with straight and dislocation-free edges and heterojunctions
(such as WS$_2$/WSe$_2$ \cite{ZhangScience17,XieScience18} and MoSe$_2$/WSe$_2$ or MoS$_2$/WS$_2$
\cite{SahooNature18} monolayer superlattices or WS$_2$/MoS$_2$/WS$_2$ in-plane multijunctions
\cite{KobayashiACSNano19}).

\begin{figure*}
  \centerline{\includegraphics[width=\textwidth]{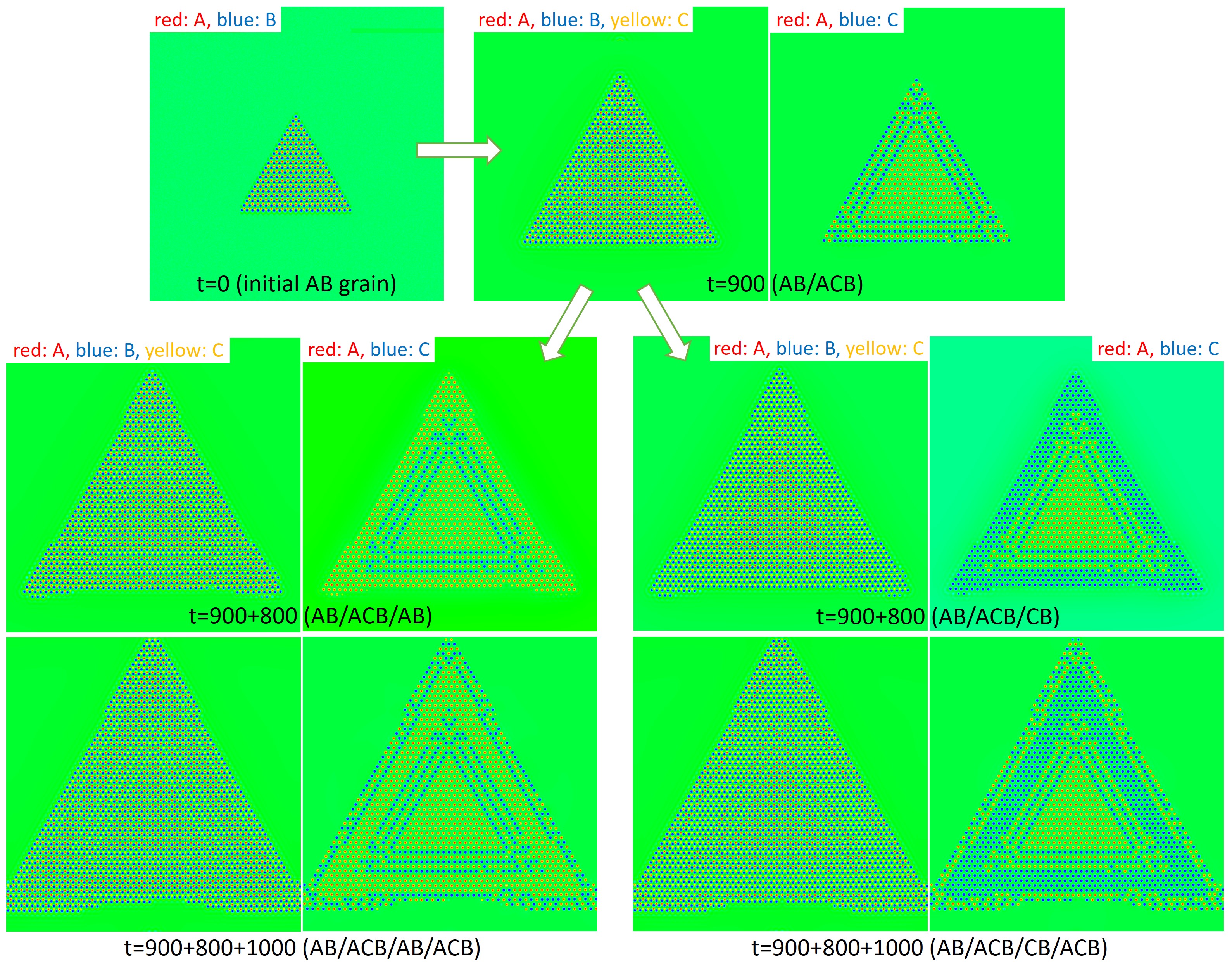}}
  \caption{The predicted growth procedure of lateral heterostructures and multijunctions
    consisting of $AB$ and $CB$ binary layers and $ACB$ ternary layers with $A$-$C$ atomic ordering.
    Both spatial configurations of $A$-$B$-$C$ atomic densities and the corresponding $A$-$C$
    structures are shown. Results are obtained from PFC simulations at $\epsilon=0.3$ and
    $\beta_{AC}=0.9$.}
  \label{fig:hetero_ABACB}
\end{figure*}

A potentially important but rarely explored configuration is a monolayer lateral heterostructure
composed of distinct blocks of 2D ternary alloys, which could bring in an additional degree of
freedom for heterostructural functionality control, in terms of the added flexibility of compositional
variation and the composition-enabled tuning of electronic properties (e.g., band gap engineering)
in each alloy block. Recently three-junction MoS$_{2(1-x)}$Se$_{2x}$/WS$_{2(1-x')}$Se$_{2x'}$ lateral
alloy heterostructures were fabricated \cite{SahooNature18}, where S and Se components in
each individual block were observed to be completely miscible, with uniform alloying. Here
we explore a different type of ternary-alloy-based lateral heterostructure or multijunction,
which integrates $ACB$-type ternary alloy domains with atomic ordering that is achievable at
low enough growth temperature, based on our related single-domain modeling in the above
Secs.~\ref{sec:order_sep} and \ref{sec:intermixing} as well as a recent experimental finding
\cite{AziziPRL20} which demonstrated the important effect of lattice atomic order on
the electronic structure of 2D ternary TMD alloy Re$_{0.5}$Nb$_{0.5}$S$_2$.

Some sample configurations of such alloy-based lateral heterostructures predicted from PFC
simulations are shown in Fig.~\ref{fig:hetero_ABACB}. Similarly, we start from the same pre-grown
$AB$ single crystal at $t=0$ as in Fig.~\ref{fig:hetero_ABCB}, which is surrounded by a homogeneous
media, but with a larger flux and higher density of $A$ component (same as that of component $C$,
e.g., $n_{A0}=n_{C0}=-0.42$ and $n_{B0}=-0.375$) outside the initial $AB$ grain. This leads to the
formation of atomically ordered $ACB$ ternary alloy from the zigzag lattice front, via the process
of lateral edge-epitaxy. The corresponding ternary lattice structure is the same as that given
in Fig.~\ref{fig:phasesep-ordering} at $\beta_{AC}=0.9$ and $\epsilon=0.3$, showing as $AB$ and
$CB$ binary honeycomb rings plus the ordering between $A$ and $C$ components (see also the
$A$-$C$ density distribution in the top-right panel of Fig.~\ref{fig:hetero_ABACB}). The
resulting $AB/ACB$ ordered heterostructure can maintain sharp heterointerface and faceted
outer surface at low enough temperature.

A similar process of sequential growth via flux control can be adopted to produce a variety of
lateral multijunctions or superlattices integrating domains of ternary 2D ordered alloys.
The periodicity of these multi-heterostructures and the types of their constituent blocks or units
could be varied and controlled. Two such examples, $AB/ACB/AB/ACB$ superlattice and $AB/ACB/CB/ACB$
multijunctions, are demonstrated in Fig.~\ref{fig:hetero_ABACB} as obtained from our PFC simulations.
For the first one, the abovementioned $AB/ACB$ growth procedure is continued but with the reduction
of $C$-species density (to $n_{C0}=-0.58$ to turn off the $CB$ growth) and the increase of $A$ flux
(with $n_{A0}=n_{B0}=-0.375$) outside the as-grown $AB/ACB$ grain, yielding the edge-epitaxy of the
next $AB$ layer and the in-plane $AB/ACB/AB$ heterostructure; switching back to the earlier flux
condition (i.e., $n_{A0}=n_{C0}=-0.42$ and $n_{B0}=-0.375$) leads to the lateral coherent growth
of the $ACB$ ordered alloy again and hence an $AB/ACB/AB/ACB$ ordered-alloy-based superlattice.
The same process of lateral epitaxy is followed in the second example to grow the in-plane
$AB/ACB/CB/ACB$ multi-heterostructure, with the only difference being in the mid step to form
the $CB$ (instead of $AB$) block from the as-grown $ACB$ alloy edge, which then requires the
setup of low enough $A$-precursor density ($n_{A0}=-0.58$) and high enough $C$ and $B$ densities
(with $n_{C0}=n_{B0}=-0.375$) in the initial homogeneous phase surrounding the $AB/ACB$ grain to
enable the subsequent formation of $AB/ACB/CB$ multijunction.

An even richer variety of lateral multi-heterostructures can be achieved by tuning the composition
of different ternary alloy blocks as well as combining ordered and disordered alloy domains with
varying types of heterointerfaces and junctions, which would result in functionally distinct but
tunable heterostructural systems particularly in terms of optoelectronic or transport properties
(with e.g., various types of spatial modulation of bandgaps and band alignments). All these
in-plane structures can be prepared via the similar edge-epitaxial growth procedure described
above with the control of deposition fluxes and temperature, which would significantly expand
the range of potential 2D material systems or configurations with controllable functionality.

\section{Conclusions}

We have developed a ternary phase field crystal model to study the growth and evolution processes
of 2D crystals and in-plane lateral heterostructures of ternary hexagonal materials that are
spatially and compositionally modulated. In this $A$-$B$-$C$ ternary alloy with each of the
$AB$ and $CB$ compounds forming a binary honeycomb structure, a transition from the phase-separated
state between $AB$ and $CB$ domains, to an $A$-$C$ atomically ordered phase with geometric
frustration as found in a recent experiment of 2D ternary TMD alloy, can be achieved by controlling
the degree of energetic preference of heteroelemental neighboring between $A$ and $C$ components.
The results are quantified through the calculations of spatial correlation functions, and
also of an intermixing factor and an overlap factor for identifying the degree of compositional
intermixing or disordering which is shown to increase at higher growth temperature, consistent
with experimental observations.

These findings for 2D single-crystalline ternary grains are used as the basics to probe the growth
of a variety of in-plane heterostructures, multijunctions, and superlattices. Sample results of
$AB/CB$, $AB/CB/AB$, and $AB/CB/AB/CB$ type multi-heterostructures are produced in our PFC
simulations through a sequential growth process of lateral edge-epitaxy, well agreeing with
recent experimental outcomes of 2D TMD/TMD lateral heterostructures and superlattices.
Importantly, our findings are extended to predict a new type of alloy-based in-plane heterostructures
integrating blocks of $ACB$-type ternary alloy with atomic ordering, such as $AB/ACB$ type ordered-alloy
heterostructure, $AB/ACB/AB/ACB$ superlattice, and $AB/ACB/CB/ACB$ multijunctions. This can be
achieved through the control of the constituent densities and fluxes at each growth stage and
the growth temperature, giving a viable way for exploring a wider variety of 2D heterostructural
material systems.

\begin{acknowledgments}
  This work was supported by the National Science Foundation under Grant No. DMR-2006446.
\end{acknowledgments}

\bibliography{tpfc_2Dhon_references}

\end{document}